\documentclass[aps,12pt,final]{revtex4b5}
\usepackage{graphicx}
\usepackage{dcolumn}
\usepackage{amsmath}

\begin{document}
\bibliographystyle{apsrev}

\title{Fractions of proton, helium, middle and heavy nuclei in primary
cosmic rays at energy $10^{16}$~eVFractions of proton, helium,
middle and heavy nuclei in primary cosmic rays at energy
$10^{16}$~eV.}

\author{Maia Kalmakhelidze, Nina Roinishvili, Manana Svanidze}
\email{ninarich@iph.hepi.edu.ge} \affiliation{E.Andronikashvili
Institute of Physics, Academy of Sciences of the Georgian
Republic, Tamarashvili 6, Tbilisi, Georgian Republic,\ 380077.}
\date{\today}

\begin{abstract}
Classification of $\gamma$ - hadron  families, registered by the
Pamir collaboration, on four groups of nuclei (P, He, middle and
heavy), responsible for their generation, is made, and fractions
of families in each of the groups are estimated. Results show,
that below the knee of the energy spectrum the chemical
composition of primary cosmic  rays remains close to the normal
one.
\end{abstract}

\maketitle

\begin{center}
\chapter {\bf1.~Introduction}
\end{center}
The interest to research of chemical composition (CC) of primary
cosmic ray (PCR) in energy area close to the knee of the energy
spectrum ($3\times{10^{15}}$ eV) does not require a reason and
confirmation. It is enough to remind recent results of
experiments: DICE/CASA-MIA \cite{1} and HEGRA \cite{2}, that
affirm essential fraction of light elements (~70 $\%$ P+He) or at
least an absence of significant changes of CC; CASA-BLANKA
\cite{3} and KASCADE \cite{4}, that observe thin structure in
dependence of $\langle \ln A\rangle $ on $E_{o}$ (a dip in knee
area) and, at last, KASCADE \cite{5} and CASA-MIA \cite{6},
declined to heavier mass composition of PCR in the knee region.

Alternative methods of PCR chemical composition research in the
same energy area, based on data obtained by means of X-ray
emulsion chambers (EC), do not find out a growth of a fraction of
heavy elements (Fe) in PCR composition \cite{7,8}. Only the part
of heavy nuclei in PCR was studied in these and also in earlier
works \cite{9,10,11}, using results of EC.

In the present work an attempt of selection of families generated
by four different groups of nuclei (P, He, middle and heavy
nuclei) and an estimation of their fraction in PCR are made.

\begin{center}
\chapter{\bf2.~Method of research.}
\end{center}

It has been shown \cite{8}, that it is reasonable to use only
three parameters of $\gamma$-hadron families for research of
chemical composition of primary cosmic rays. They are: $n_{h}$~-~
the number of hadrons in  a  family,  $R\gamma$~-~an  average
radius of $\gamma$-component of the  family  and $d
=n_{in}/n_{obs}$~-~a ratio of "initial" to observed numbers of
$\gamma$-quanta. These
 parameters, on the one hand, are sensitive to the mass
of an induced nucleus, and on the other, do not correlate with
each other. In order to reduce the number of used parameters to
one (i.e.to turn the analysis to a one-dementional form, similar
to the neural nets method) it has been suggested \cite{8} to use
one third of a sum of normalized parameters:

\begin{equation}
X=(X_{n_h}+  X_{R_\gamma}+ X_d)/3
\end{equation}
where
\begin{equation}
X_{n_h} = (n_h - \langle n_h\rangle) / \delta{n_h},~~~
X_{R_\gamma} = (R_\gamma - \langle R_\gamma\rangle) /
\delta{R_\gamma},~~~X_{d} = (d - \langle d\rangle) / \delta{d}
\end{equation}
In [2] $\langle n_h\rangle$ , $\delta{n_h}$, $ \langle
R_\gamma\rangle$ , $\delta{R_\gamma}$ and $\langle d\rangle$ ,
$\delta{d}$  are determined on the base of artificial families
generated by protons.  It has to be mentioned, that for simulation
of families we use the MC0 model \cite{12}.

The sensitivity of X to the mass of generating nucleus is
demonstrated on Figure 1, that shows its distributions for four
groups of nuclei: P, He, middle nuclei and Fe.The group of middle
nuclei includes C, N, O, Mg and Si. The figure is drown for
simulated $\gamma$- families satisfying the same selection rules
as experimental families (see below).

The conditions used to select the families similar to those
generated by the given nucleus "P-like", "He-like", "Mid-like",
"Fe-like", are brought in Table  ~\ref{tab1}.

\begin{table}
\caption{Selection conditions of families}.\\
 \label{tab1}
\begin{tabular}{|c|c|c|}\hline
 ~~~ N of group,~$\bf{j}$~~~ &~~~ Type of families~~~ &~~~ Conditions \\\hline
  ~~~1~~~ &~~~ P-like~~~ &~~~$ X<-0.06$~~~ \\\hline
  ~~~2~~~ & ~~~He-like~~~ &~~~ $0.3<X<1.2$~~~ \\\hline
  ~~~3~~~ &~~~ Mid-like~~~ &~~~ $0.7<X<1.5$~~~ \\\hline
  ~~~4~~~ &~~~ Fe-like~~~ &~~~ $X>1.04$~~~ \\ \hline
\end{tabular}
\end{table}
The fraction of families, related to  group of nuclei $\bf{j}$ is
determined by the following:
\begin{equation}
f_{j}^{'} = \Sigma_{A}(f_j \times {R_A})
\end{equation}
where   $ f_{j}$   is  real fraction of families, generated by
nuclei of the group $\bf{j}$ , and $\bf{R_{A}}$ -is a fraction of
families from A nucleus related to the group $\bf{j}$.

If A and $\bf{j}$ coincide, it means, that the identification is
made correctly, otherwise the identification is false. To restore
the real fraction $f_{j}$ the expression [3] turns:
\begin{equation}
f_{j} =[f_{j}^{'}- \Sigma^{'}_{A}(f_j \times {R_A})]/R_A
\end{equation}
In $\Sigma^{'}$  does not include the member correspondly A=j. In
other words, $\Sigma^{'}$ summes the contributions of all falsy
identified families.

Values of  $R_{A}$ were determined using training sets of
artificial families generated by nucleus A.

 From the stated above , the suggested method of chemical
composition research is reduced to selection of groups of
families,that satisfy the conditions brought in Table~\ref{tab1},
 to estimation of their fraction $f^{'}_{j}$ and to restoration of
real fraction of families $f_{j}$.

\begin{center}
\chapter{\bf3.~Testing the method efficiency.}
\end{center}

\begin{table}

\caption{Fractions of nuclei $C_{A}$ in analyzed chemical
compositions of PCR and appropriate fractions of families from
nuclei A, $F_{A}$. $C_{A}$ and $F_{A}$  in $\%$.}
 \label{tab2}
\begin{tabular}{|c|c|c|c|c|c|c|}\hline
 ~~~ Composition~~~ &~~~
   &~~~ P~~~ &~~~He~~~ &~~~CNO~~~ &~~~SiMg ~~~&~~~ Fe\\\hline
 ~~~ Normal~~~&~~~ $C_{A}$~~~&~~~ 40~~~ &~~~20 ~~~&~~~10~~~ &~~~ 10~~~ &~~~ 20~~~ \\
  \cite{13} & $F_{A}$ & 76 & 16 & 4 & 2.5 & 2.3 \\\hline
  Superheavy &$C_{A}$ & 7 & 5 & 12 & 6 & 70 \\
  \cite{14} & $F_{A}$ & 40 & 13 & 12 & 5 & 31 \\ \hline
\end{tabular}
\end{table}

To study the efficiency of the proposed method test families, to
generate these families P, Fe and two complex chemical
compositions were used with the characteristics shown in
Table~\ref{tab2}.

True fractions of families $\bf{F_{A}}$ from nucleus A, which
should be compared with the restored fraction $f_{j}$ in the case
of complex composition is expressed by the formula:
\begin{equation}
F_{A} =\epsilon_{A}\times{C_A}/\Sigma_A(\epsilon_{A}\times{C_A})
\end{equation}
Registration efficiency $\epsilon_{A} = N_{f} / N_{A}$ , where
$N_{A}$ is the number of simulated primary nuclei, and $N_{f}$ -
 is the number of families generated by them.

 Table~\ref{tab3} compares the true values $F_{A}$ from(5) with
the restored values $f_{j}$ from(4).

\begin{table}

\caption{Fractions of nuclei in analyzed chemical compositions of
PCR, $C_{A}$, and appropriate fractions of families from nuclei A,
$F_{A}$. $C_{A}$ and $F_{A}$  in $\%$.}
 \label{tab3}
\begin{tabular}{|c|} \hline
~~~~~~~~~~~~~~~~~~~~~~~~~~~~~~~~~~~~~~~~~~~~~~~~~~~ Groups of
families~~~~~~~~~~~~~~~~~~~~~~~~~~~~~~~~~~~~~~~~~~~~~~\\ \hline
\end{tabular}
\begin{tabular}{|c|c|c|c|c|}
~~~~~~~~~~~~~~~~~~~~~~~~~&~~~~~~~~~P(j=1)~~~~~&~~~~~~~~He(j=2)~~~~~~~&~~~~~~Mid(j=3)~~~~~~
&~~~~~~Fe(j=4)~~~~~~~\\ \hline
\end{tabular}
\begin{tabular}{|c|c|c|c|c|c|c|c|c|}
~~~~~~composition~~~~&~~~~~$f_{1}$~~~~&~~~~$F_{P}$~~~&~~~~$f_{2}$~~~~~~
&~~~$F_{\alpha}$~~~~~&~~~~$f_{3}$~~~~~~&~~~~$F_{Mid}$~~&~~~~$f_{4}$~~~~
&~~~~$F_{Fe}$~~
\\ \hline
  P& 1.08$\pm$& 1.0&0.023$\pm$&0&0.017$\pm$&0&0.026$\pm$&0\\
& 0.04& &0.034&&0.015&&0.013&\\\hline
 Fe& 0.0$\pm$&0 &-0.059$\pm$&0&-0.017$\pm$&0&1.027$\pm$&1.0\\
& 0.02& &0.068&&0.04&&0.05&\\\hline
  Normal& 0.74$\pm$& 0.76&0.16$\pm$&0.155&0.033$\pm$&0.05&0.053$\pm$&0.025\\
& 0.03& &0.03&&0.015&&0.015&\\\hline
 Superheavy& 0.41$\pm$& 0.40&0.13$\pm$&0.13&0.09$\pm$&0.17&0.38$\pm$&0.31\\
& 0.03& &0.05&&0.02&&0.03&\\ \hline
\end{tabular}
\end{table}

 Good agreement between expected values  $F_{A}$ and
restored values $f_{j}$ by means of the proposed method of family
classification convinces in a rather high efficiency of the
mathod.

\begin{center}
\chapter{\bf4.~Experimental results.}
\end{center}
 The suggested method  has been applied to 174 experimental families of
 the Pamir collaboration. The results are given in Table~\ref{tab4}.
  Family satisfy the following criteria:

 $100~TeV < \Sigma{E_{\gamma}}<1000~ TeV;~~~  E_{\gamma},~ E^{\gamma}_{h} > 4 TeV;~~~   R_{\gamma} >
 1cm.$
\begin{table}
\caption{Fractions of experimental families, related to different
groups of nuclei, $f_{j}$.} \label{tab4}
\begin{tabular}{|c|c|c|c|c|}\hline
  ~~~~ &~~~ P~~~ &~~~ He~~~ &~~~Mid~~~ &~~~ Fe~~~ \\\hline
  ~~~$f_{j}$~~~ &~~~0.92$\pm$0.11~~~ &~~~ 0.05$\pm$0.10~~~ &~~~ 0.01$\pm$0.04~~~ &~~~0.04$\pm$0.04~~~ \\ \hline
\end{tabular}
\end{table}

Table~\ref{tab4} shows, that the fractions of families, generated
by light nuclei at energies below the knee of PCR energy spectrum,
remain high,while the middle and heavy nuclei within the error
limits of estimations, are present in PCR as in the normal CC.

\begin{center}
\chapter{\bf5.~Conclusions.}
\end{center}
In the present work we show that:

1. $\gamma$-hadron families allow  their rather effective
classification on groups of nuclei responsible for their
generation and, hence, they can be used for successful
investigation of chemical composition of primary cosmic rays.

2. Analysis of $\gamma$-hadron families brings to the conclusion
that the chemical composition of primary cosmic radiation below
the knee of its energy spectrum remains the normal or is displaced
towards the light nuclei.


\begin{figure}
\includegraphics{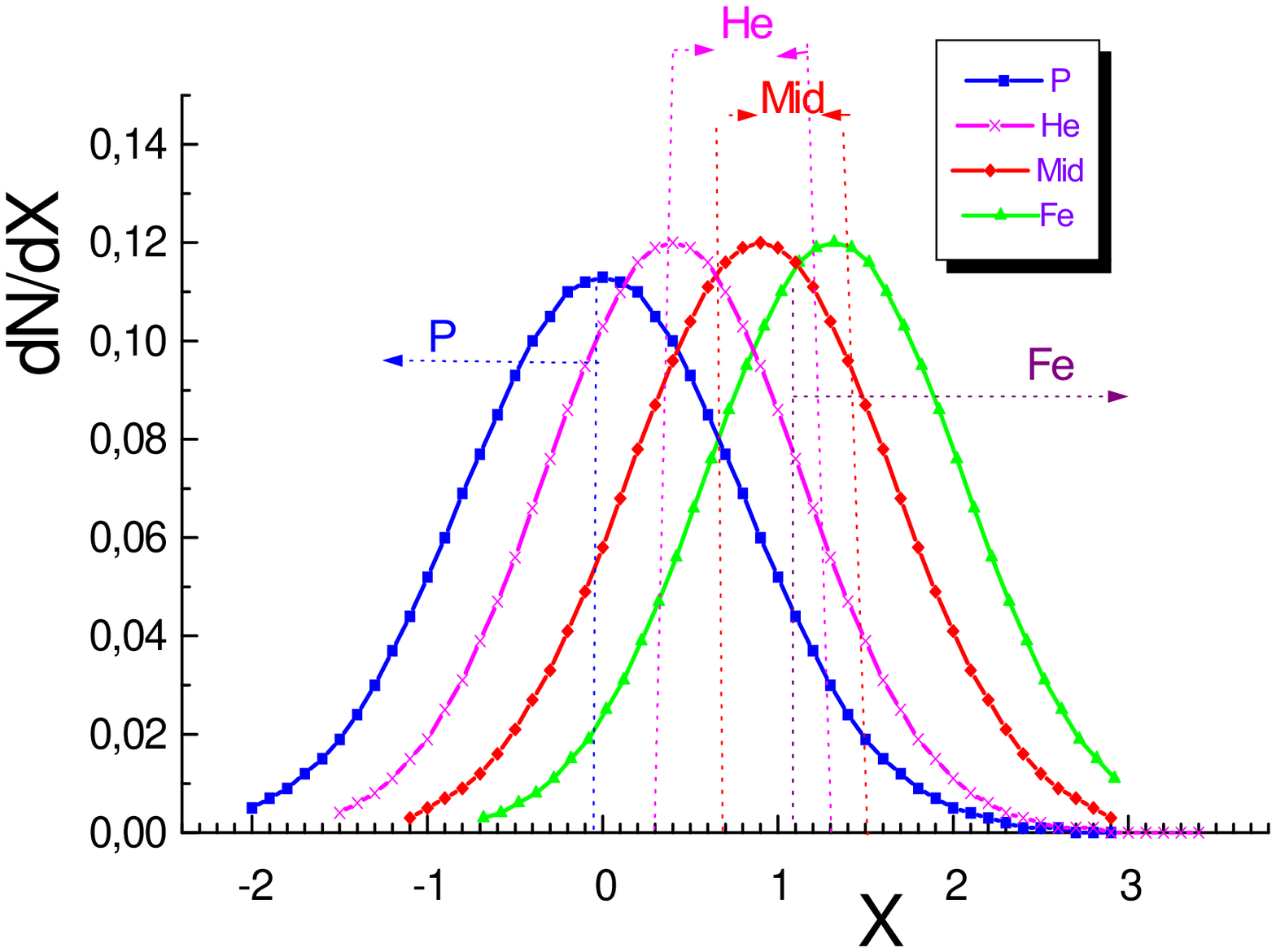}
\caption{Distributions of parameter X  for  four groups of nuclei.
From left to the right :P, He, middle and heavy nuclei. The arrows
correspond to conditions of classification brought in
Table~\ref{tab1}.}
\end{figure}

\end{document}